\pdfoutput=1

\documentclass[11pt,twoside,a4paper,cmspaper,final,collab]{cms-tdr}
\def\svnVersion{dd8d592-D}\def\svnDate{2019/12/03}\def\cmsCernNoTag{CERN-EP-2019-068}\def\cmsCernDate{\today}\def\cmsMessage{"Published in Physical Review Letters as \href{http://dx.doi.org/10.1103/PhysRevLett.123.231803}{\doi{10.1103/PhysRevLett.123.231803}.}"}

\begin{document}\cmsNoteHeader{EXO-17-027}

\hyphenation{had-ron-i-za-tion}
\hyphenation{cal-or-i-me-ter}
\hyphenation{de-vices}
\RCS$HeadURL$
\RCS$Id$

\newcommand{\gpq}{\ensuremath{g^\prime_\cPq}}
\newcommand{\msd}{\ensuremath{m_{\mathrm{SD}}}}
\newcommand{\nddt}{\ensuremath{N^2_{\mathrm{DDT}}}}

\newlength\cmsFigWidth
\ifthenelse{\boolean{cms@external}}{\setlength\cmsFigWidth{0.49\textwidth}}{\setlength\cmsFigWidth{0.9\textwidth}}
\ifthenelse{\boolean{cms@external}}{\providecommand{\cmsLeft}{upper\xspace}}{\providecommand{\cmsLeft}{left\xspace}}
\ifthenelse{\boolean{cms@external}}{\providecommand{\cmsRight}{lower\xspace}}{\providecommand{\cmsRight}{right\xspace}}
\ifthenelse{\boolean{cms@external}}{\providecommand{\NP}{\ensuremath{N_P}\xspace}}{\providecommand{\NP}{\ensuremath{N_\mathrm{P}}\xspace}}
\ifthenelse{\boolean{cms@external}}{\providecommand{\NF}{\ensuremath{N_F}\xspace}}{\providecommand{\NF}{\ensuremath{N_\mathrm{F}}\xspace}}
\ifthenelse{\boolean{cms@external}}{\providecommand{\CLnp}{\ensuremath{\text{C.L}}\xspace}}{\providecommand{\CLnp}{\ensuremath{\text{CL}}\xspace}}
\ifthenelse{\boolean{cms@external}}{\providecommand{\CL}{\ensuremath{\text{C.L.}}\xspace}}{\providecommand{\CL}{\ensuremath{\text{CL}}\xspace}}

\cmsNoteHeader{EXO-17-027}
\title{Search for low-mass quark-antiquark resonances produced in association with a photon at \texorpdfstring{$\sqrt{s} = 13$\TeV}{sqrt(s) = 13 TeV}}

\date{\today}

\abstract{
A search for narrow low-mass resonances decaying to quark-antiquark pairs is presented. The search is based on proton-proton collision events collected at 13\TeV by the CMS detector at the CERN LHC. The data sample corresponds to an integrated luminosity of 35.9\fbinv, recorded in 2016. The search considers the case where the resonance has high transverse momentum due to initial-state radiation of a hard photon. To study this process, the decay products of the resonance are reconstructed as a single large-radius jet with two-pronged substructure. The signal would be identified as a localized excess in the jet invariant mass spectrum. No evidence for such a resonance is observed in the mass range 10 to 125\GeV. Upper limits at the 95\% confidence level are set on the coupling strength of resonances decaying to quark pairs. The results obtained with this photon trigger strategy provide the first direct constraints on quark-antiquark resonance masses below 50\GeV obtained at a hadron collider.}

\hypersetup{
pdfauthor={CMS Collaboration},
pdftitle={Search for low-mass quark-antiquark resonances produced in association with a photon at sqrt(s) = 13 TeV},
pdfsubject={CMS},
pdfkeywords={CMS, dijets, resonances, photons}}

\maketitle
New resonances coupling to pairs of quarks (generally referred to as \PZpr) are ubiquitous signatures in theories beyond the standard model (SM), appearing in dark matter models~\cite{Abercrombie:2015wmb, Boveia:2016mrp} and models with extra dimensions~\cite{ref_rsg}, amongst others~\cite{Anchordoqui:2008di, Cullen:2000ef, ref_axi, Chivukula:2011ng, ref_coloron, Baur:1989kv}. The first dijet searches at a hadron collider were performed by UA1~\cite{Albajar:1988rs} and UA2~\cite{Alitti:1993pn}, and have been extended to higher resonance masses by CDF~\cite{Aaltonen:2008dn} and D0~\cite{Abazov:2003tj} at the Tevatron, and by ATLAS~\cite{PhysRevD.96.052004} and CMS~\cite{Sirunyan:2018xlo} at the LHC. However, as collision energy and beam intensity have increased, there has been a loss of sensitivity to lower mass resonances, which stems from the increasing cross section of background multijet events, tighter online requirements needed to handle growing event rates, and the large numbers of simultaneous collisions per bunch crossing (pileup). These issues can be partially mitigated by focusing on events in which the resonance is produced in association with high momentum initial-state radiation (ISR). In such a scenario, the two quarks hadronize into a single massive jet. In particular, by considering events with a high transverse momentum (\pt) ISR photon or jet, the ATLAS Collaboration searched for a \PZpr decaying to quark-antiquark pairs~\cite{Aaboud:2019zxd} and reported a result for resonance masses as low as 100\GeV. The CMS Collaboration used this method with ISR jets to search for \PZpr with masses as low as 50\GeV~\cite{Nhan:2017aaa}, the lowest mass then probed by collider experiments.

This analysis, which considers events produced with ISR photons from pp collisions at $\sqrt{s} = 13$\TeV, using data collected by the CMS detector in 2016, and corresponding to an integrated luminosity of 35.9\fbinv, extends dijet searches to low \PZpr masses where only indirect measurements~\cite{Dobrescu:2014fca} provide constraints on the hadronic production of such new physics. This extension to low \PZpr masses is possible in this analysis because of the reliance on a photon trigger, for which it is feasible to select dijet events using a lower \pt threshold than for jet triggers. However, the mass of the \PZpr is sufficiently low compared to its momentum that the separate hadronizations of the resulting quark and antiquark merge into a single large-radius jet. This search is performed by looking for a localized excess in the jet mass spectrum in events with a photon and a jet with the two-pronged jet substructure expected for the signal.

The main background, arising from photons produced in association with jets by SM processes, is derived using a data-driven method. Additional resonant SM background processes, composed of \ttbar events and the SM production of $\PW+\gamma$ and $\PZ+\gamma$, are estimated from simulation, with corrections obtained from control regions in data. The results are interpreted within the framework of a \PZpr with mass between 10 and 125\GeV, decaying into quarks, and are used to set limits on the quark coupling $\gpq$ as a function of the \PZpr mass.

The CMS detector consists of a silicon tracker, a lead tungstate crystal electromagnetic calorimeter (ECAL), a brass and scintillator hadron calorimeter (HCAL), and gas-ionization muon detectors. A superconducting solenoid provides a uniform magnetic field within the detector. Events are sorted by a two-tiered triggering system~\cite{Khachatryan:2016bia} to ensure that only events of potential physics interest are recorded. A more detailed description of the CMS detector, including its angular coordinates $\eta$ and $\phi$, can be found in Ref.~\cite{Chatrchyan:2008zzk}.

Events are reconstructed using the CMS particle-flow (PF)~\cite{Sirunyan:2017ulk} algorithm, which combines information from every element of the CMS detector to reconstruct and identify individual particles (called PF candidates). Each particle is classified as either a muon, electron, photon, charged hadron, or neutral hadron. The energy of photons is obtained directly from ECAL measurements. Similar measurements, along with information from the tracker, are used to determine the energy of electrons. Misidentification of particles is possible, so additional isolation and purity requirements on potential photons are imposed~\cite{Khachatryan:2015iwa}. The momentum of muons is measured from the curvature of their tracks. Neutral and charged hadron energies are measured from their deposits in the ECAL and HCAL, with information from the tracker used to further constrain the energy of the charged hadrons. The missing transverse momentum (\ptvecmiss) is defined as the negative vector \pt sum of all reconstructed particles in an event. The PF candidates are clustered into jets using FastJet~\cite{Cacciari:2011ma} with the anti-\kt algorithm~\cite{Cacciari:2008gp} and a distance parameter of 0.4 and 0.8 for AK4 and AK8 jets, respectively. Particles produced in additional collisions within the same bunch crossing are suppressed by applying a weight to each PF candidate, calculated by the pileup-per-particle identification~\cite{Bertolini:2014bba} algorithm. Jets are corrected as a function of their $\pt$ and pseudorapidity ($\eta$) to match the observed detector response~\cite{jec}. Jets arising from the hadronization of \Pb quarks are identified using the CSVv2 algorithm~\cite{Sirunyan:2017ezt}.

The signal benchmark model~\cite{PhysRevD.90.079901} used in this analysis and in Refs.~\cite{Aaboud:2019zxd} and~\cite{Nhan:2017aaa} features a vector resonance \PZpr, with the coupling constant to quarks set to $\gpq=1/6$, at which the \PZpr width is well below the resolution of the detector. It was simulated to leading order with the \MGvATNLO~\cite{Alwall:2014hca} generator, with MLM matching~\cite{Alwall:2007fs} between jets from matrix element calculations and the parton showers. Up to 3 additional jets are allowed in the matrix element calculation. The model assumes no interaction between the SM \PZ and the \PZpr. The same generator is used to model at leading order the quantum chromodynamic production of multijet events, which can include radiated photons, and the $\gamma$+jets background, where the photon is part of the hard interaction, as well as, to next-to-leading order, the backgrounds $\PW+\gamma$ and $\PZ+\gamma$. The multijet and $\gamma$+jet components are treated together as a single nonresonant background, with the angle between the leading photon and the nearest jet used to define a phase space for each sample. Events from the multijet sample are removed if they are in the $\gamma$+jets phase space. The \POWHEG 2.0~\cite{POW1,POW2,POW3} generator is used to model \ttbar events at next-to-leading order. All signal and background generators are interfaced with \PYTHIA\,8.212~\cite{Sjostrand:2014zea}, with the \textsc{CUETP8M1} underlying event tune~\cite{Khachatryan:2015pea}, to simulate parton showering and hadronization effects. The generated events are processed through a \GEANTfour~\cite{GEANT4} simulation of the CMS detector. This simulation includes effects from both in-time and out-of-time pileup. The parton distribution function set NNPDF3.0~\cite{Ball:2014uwa} is used to produce all simulated samples. Where necessary, differences between the reconstruction of simulated and real quantities are corrected by applying scale factors to the simulation, derived from control regions in data~\cite{jec}.

The trigger strategy used by this search is to require one photon with $\pt>175\GeV$ and $\abs{\eta}<3.0$. To ensure a full triggering efficiency for events that satisfy the subsequent selection, offline photons are required to have $\pt>200\GeV$ and $\abs{\eta}<2.4$. Events with additional identified photons of $\pt>14\GeV$ or leptons of $\pt>10\GeV$ are discarded to avoid overlap with other searches and to reduce backgrounds from electroweak sources. Even leptons in a pair that are sufficiently co-linear to be reconstructed as a single jet are generally also tagged as separate leptons and thus excluded. The $\PZpr\to\qqbar$ decay is assumed to correspond to the highest momentum AK8 jet in the event. Only events with leading jet $\pt>200\GeV$ are considered. To reduce the contribution from \ttbar, events with an AK4 jet with $\pt>30\GeV$ and satisfying the loose working point~\cite{Sirunyan:2017ezt} of the CSVv2 algorithm (excluding AK4 jets within $\Delta R<0.6$ of the leading AK8 jet), or with $\ptmiss>75\GeV$, are discarded. A separation $\sqrt{\smash[b]{(\Delta\eta)^2+(\Delta\phi)^2}} > 2.2$ is required between the leading AK8 jet and the photon in the event.

The soft drop mass algorithm (with $\beta=0, z_{\text{cut}}=0.1$) ~\cite{Dasgupta:2013ihk,Larkoski:2014wba} is used to remove soft and wide-angle radiation from the jet, and the resulting distribution of ``groomed jet mass'' ($\msd$) is inspected for localized excesses. The modeling of $\msd$ has been tested for masses only down to 10\GeV~\cite{JMASSPROOF}; thus 10\GeV is the lowest signal mass considered by this analysis. The highest signal mass considered is 125\GeV, above which there is a low probability of reconstructing the \PZpr as a single jet. The selected events are divided into signal and control regions based on the $\eta$ of the photon, with the boundary between the regions chosen to maximize the sensitivity of the analysis. Events with photon $\abs{\eta}<2.1$ are considered to be in the signal region. The events with $\abs{\eta}>2.1$, in which the photon is more likely to have been radiated in a multijet process rather than in a hard scattering, define the $\eta$ control region to perform substructure measurements of jets with kinematic variables similar to those of jets in the signal region. These variables are computed only on jet constituents that have survived the soft drop algorithm.

The variable $N_2^1$~\cite{Larkoski:2013eya,Moult:2016cvt} is used to further separate signal jets with two-pronged substructure from the background. This variable is defined using a combination of functions that correlate angles among the constituents of the jet to categorize the substructure. A jet originating from a two-pronged decay is more likely to have a low value of $N_2^1$. In addition, we define the dimensionless quantity $\rho = \ln({m^2}/{\pt^2})$~\cite{Dolen:2016kst}, which, unlike the mass itself, is approximately uncorrelated with the jet \pt.

While the $N_2^1$ variable offers a considerable discrimination power, the background efficiency for retaining jets based on a fixed cut on $N_2^1$ has dependencies on the jet $\rho$ and $\pt$. These lead to distortion of the $\msd$ distribution, making a search for a peak difficult. To preserve the shape of the mass distribution, a varying cut on $N_2^1$ is used to remove 90$\%$ of the background. To achieve this, a decorrelated variable $\nddt$ is built, which is similar to the one proposed in Ref.~\cite{Dolen:2016kst}. This variable is defined as:
\begin{equation}
    N_2^{\mathrm{DDT}} (\rho^{\text{jet}}, \pt^{\text{jet}}) = N_2^1 - X_{10\%} (\rho^{\text{jet}}, \pt^{\text{jet}}),
\end{equation}
where $X_{10\%}$ is the value of $N_2^1$ where a cut would retain 10\% of the background. The values for $X_{10\%}$ in bins of the jet $\rho$ and $\pt$ are taken from the $\eta$ control region. A smoothing procedure is applied to the $X_{10\%}$ distribution to reduce unphysical features where the statistical uncertainty is large. The selection $\nddt<0$ is applied to events in the signal region. By construction, this selection will have a background efficiency of exactly $10\%$ for the sample in which $\nddt$ was constructed. Signal hypotheses across the entire parameter range of the analysis were injected into simulated background distributions to evaluate potential contamination of the $\eta$ control region and its effect on the mass distribution of events passing the $\nddt<0$ requirement. The effects were found to be negligible compared to the statistical uncertainty in the mass distribution. Differences in the \pt and $\rho$ dependence of $X_{10\%}$ in the signal and control regions are expected, and are explicitly parametrized as part of the background estimation procedure.

The dominant background is due to nonresonant events in which a light quark or gluon jet passes the $\nddt$ requirement. The second component consists of events with two-pronged jets arising from a mixture of $\PZ+\gamma$, $\PW+\gamma$ and \ttbar events. Events from \ttbar production enter the signal selection largely through electrons being incorrectly reconstructed as photons. Other sources of background were found to be negligible. The nonresonant background is estimated from a data-driven method described below, with the simulated samples used for validation only. The other backgrounds are taken from simulation and their shapes and normalizations allowed to vary in a final fit of the passing and failing regions, with a correction derived from a \ttbar control region, also described below.

The nonresonant background in the signal region is estimated by considering the events that have passed all selection requirements except $\nddt<0$. In the $\eta$ control region, the pass-to-fail ratio of background events for the $\nddt<0$ requirement is one to nine, independent of jet \pt and $\rho$. In the signal region, this ratio is taken to be a smooth function $\mathcal{F}$($\pt,\rho$) which models the differences between the $\nddt$ variable in the $\eta$ control and signal regions. The relationship between passing (\NP) and failing (\NF) events is then: $\NP = \mathcal{F}(\pt,\rho)\NF$, for each bin of \pt and $\rho$. The deviation of $\mathcal{F}$ from a flat ratio corresponds to the difference between the signal and control regions. The unknown function $\mathcal{F}$ is expanded into a polynomial series:
\begin{eqnarray}
	\mathcal{F}(\pt,\rho) = \sum_{i}\sum_j a_{ij}\pt^j \rho^{i},
\end{eqnarray}
where the unknown coefficients $a_{ij}$ are determined by a simultaneous likelihood fit of the passing and failing events, in which the signal and resonant backgrounds are allowed to float. The number of coefficients in the fit is determined by performing the Fisher F-Tests~\cite{ref:ftest} on progressively higher order polynomial combinations of \pt and $\rho$. The optimal polynomial form is found to be third order in both \pt and $\rho$.

While this fit to data ensures that differences in the nonresonant background modeling of the $\nddt$ variable are accounted for, consistent behavior in data and simulation for resonances is not assured. A dedicated \ttbar control region is defined, built from events containing a high-\pt muon, with the selection optimized to be dominated by \ttbar production. The efficiency of the $\nddt<0$ requirement is measured by fitting the $\PW$-mass peak (where the hadronization products from both quarks merge into one jet) in the passing and failing jet mass distributions of this control region, for both data and simulated samples. This efficiency, an explicit parameter of the fit, is used to correct relative yields for resonant \ttbar events and the $\PW+\gamma$ and $\PZ+\gamma$ backgrounds obtained from simulation in the passing and failing regions. The data-to-simulation efficiency scale factor is found to be $0.909\pm0.046$ (stat+sys), and is applied to all the resonant backgrounds, as well as to the signal.

\begin{figure}[htb]
  \centering
    \includegraphics[width=\cmsFigWidth]{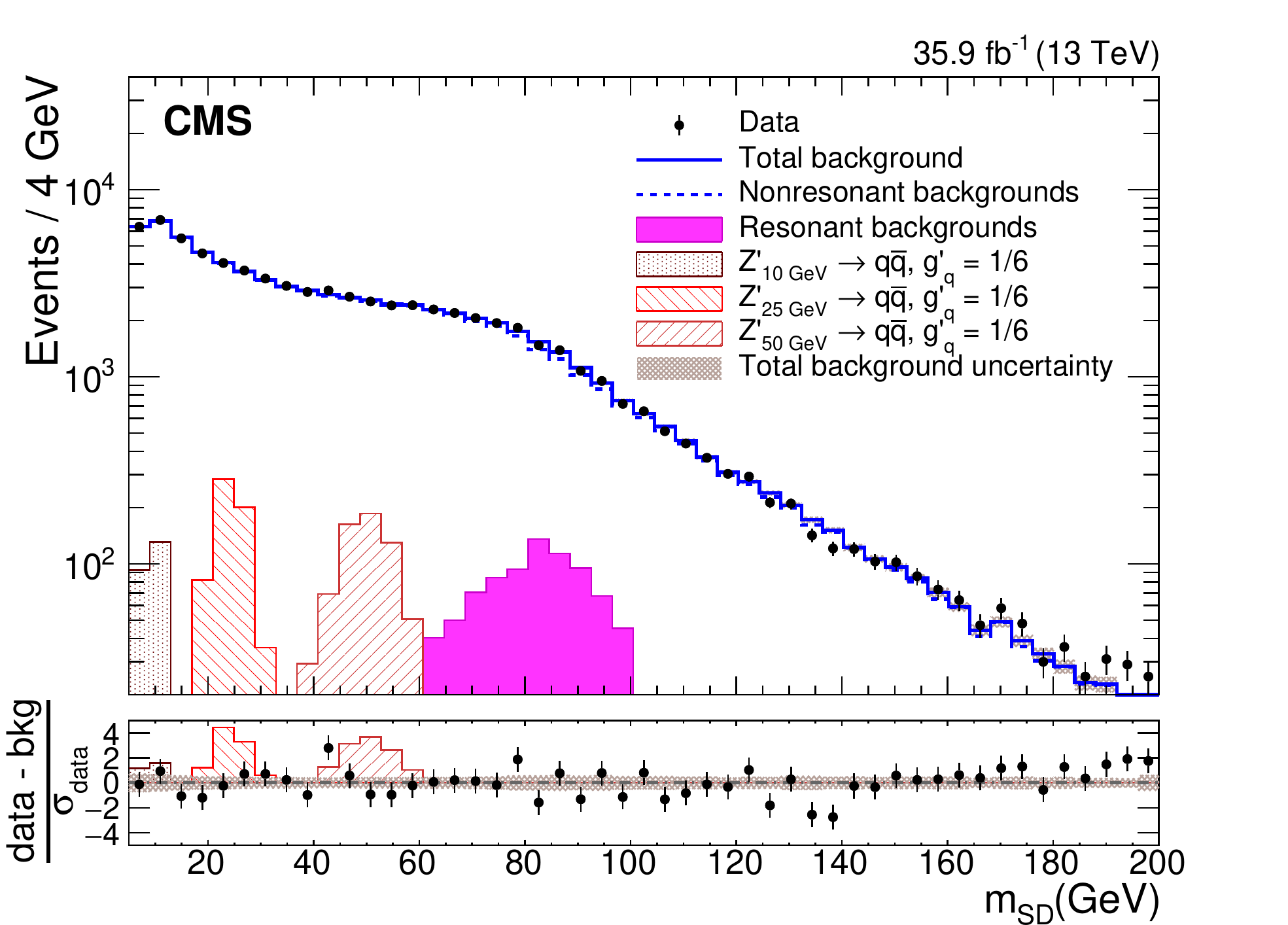}
    \caption{The soft drop~\cite{Dasgupta:2013ihk,Larkoski:2014wba} jet mass distribution of the signal region after the main background estimation fit is performed. The nonresonant background is indicated by a dashed line, while the total background composed of the sum of this nonresonant background and the resonant backgrounds is shown by the solid line. Representative signals are plotted for comparison. The bottom panel shows the difference between the data and the final background estimate, divided by the statistical uncertainty of the data in each bin. The shaded region represents the total uncertainty in the background estimate in each bin.}\label{SR}
\end{figure}

To model the $\msd$ distribution in the signal region, a binned 2D maximum likelihood fit is performed on the events passing and failing the $\nddt<0$ requirement, in all (\pt,$\rho$) bins of the signal region~\cite{Nhan:2017aaa}. In the fit, all SM processes and the signal are allowed to float simultaneously. Signal shapes are taken from simulation. The fit is performed for the background-only (null) hypothesis and for signal hypotheses for each simulated signal mass (10, 25, 50, 75, 100 and 125\GeV), as well as for interpolated mass shapes derived by vertical template morphing~\cite{READ1999357} these simulated event distributions to cover a signal hypothesis in steps of 5\GeV from 10 to 125\GeV. To ensure proper modeling of the high mass tail, the fit is performed on events with masses up to 201\GeV. The $\msd$ distribution of the signal region, summed over all \pt and $\rho$ bins, is shown in Fig.~\ref{SR}. The contributions from resonant backgrounds are evaluated as part of the likelihood, with their shapes and normalizations allowed to vary within the systematic uncertainties in the initial estimates (see Table \ref{SysTable}). The average value of the nonresonant background efficiency in the signal region determined by the fit is 9\%.

The uncertainty in the nonresonant background originates from the systematic uncertainty in the fit and the statistical uncertainty from the number of events in the region failing the $\nddt<0$ requirement. The signal, \ttbar, $\PW+\gamma$ and $\PZ+\gamma$ backgrounds are affected by correlated shape and normalization uncertainties. We constrain the efficiency of the selection based on $\nddt$ in the \ttbar control region, with the scale factor uncertainty applied to the yields of signal and the resonant backgrounds in the final fit to the signal region. The jet mass scale and resolution uncertainties are considered as uncertainties in the shape of the signal and the resonant background components in the fit. Finally, uncertainties associated with the jet energy corrections~\cite{jec}, trigger efficiency, lepton veto efficiency, resonant background normalizations and the integrated luminosity determination~\cite{lumi} are applied to the expected yields of the signal and the resonance backgrounds. These are summarized in Table \ref{SysTable}.
\renewcommand{\arraystretch}{1.25}
\begin{table}[htb]
\centering
\caption{The systematic uncertainties included in the computation of the limit on the coupling strength of \PZpr to quarks. Parameters denoted by the $\star$ symbol affect both the shape and normalization of the affected processes; otherwise only the normalization is modified. The parameters affecting normalizations have log-normal priors, and those affecting the shape have Gaussian priors, unless marked with the $\dagger$ symbol, which denotes that this parameter was floating and constrained by the final simultaneous fit of the passing and failing distributions.}
\begin{scotch}{l c c}
Systematic effect & Affected processes & Uncertainty (\%) \\
\hline
Polynomial fit$^{\dagger\star}$ & Nonresonant & 1--5\\
Electron veto & \ttbar, $\PW$, $\PZ$, \PZpr & 0.5 \\
Muon veto & \ttbar, $\PW$, $\PZ$, \PZpr & 0.5 \\
Jet mass smear$^{\dagger\star}$ & \ttbar, $\PW$, $\PZ$, \PZpr & 0.7 \\
Jet energy corrections & \ttbar, $\PW$, $\PZ$, \PZpr & 2 \\
Luminosity & \ttbar, $\PW$, $\PZ$, \PZpr & 2.5 \\
Trigger$^\star$ & \ttbar, $\PW$, $\PZ$, \PZpr & 3 \\
$\nddt$ efficiency & \ttbar, $\PW$, $\PZ$, \PZpr & 5 \\
Photon ID & \ttbar, $\PW$, $\PZ$, \PZpr & 6 \\
Jet mass scale$^{\dagger\star}$ & \ttbar, $\PW$, $\PZ$, \PZpr & 6 \\
$\PW+\gamma$ normalization$^\dagger$ & $\PW$ & 11 \\
$\PZ+\gamma$ normalization$^\dagger$ & $\PZ$ & 45 \\
\ttbar normalization$^\dagger$ & \ttbar & 54 \\
\end{scotch}
\label{SysTable}
\end{table}
To validate the robustness of the fit, a goodness-of-fit test and bias tests are performed using simulated data with a variety of simulated signals injected. No significant bias is observed for any \PZpr mass.

\begin{figure}[htb]
  \centering
   \includegraphics[width=\cmsFigWidth]{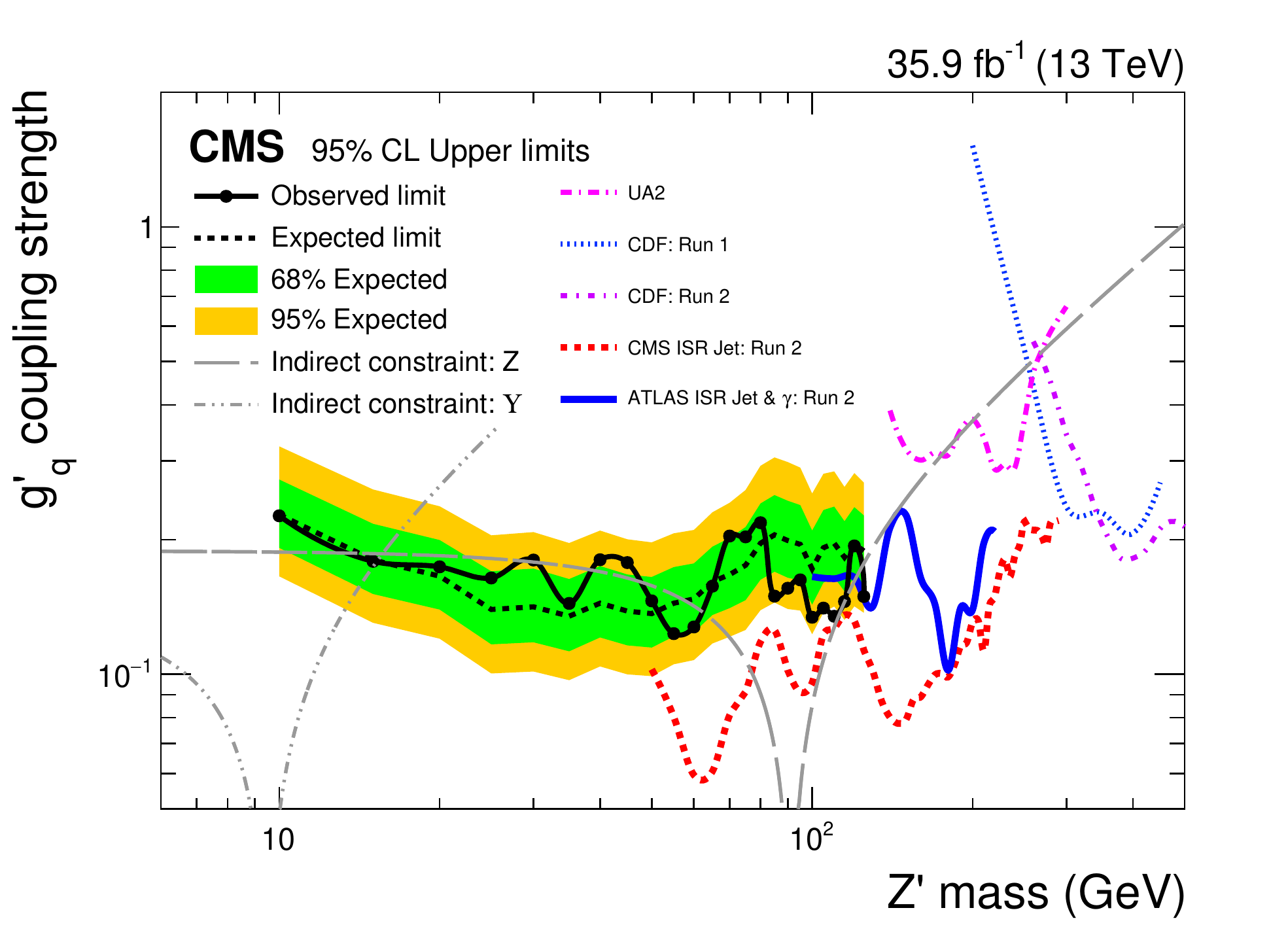}
    \caption{Upper limits at 95\% \CL on the coupling strength $\gpq$ of $\PZpr\to\qqbar$. The observed limit is shown as a solid black line, while the expected limit is dashed. The green (dark) and yellow (light) bands represent 1 and 2 standard deviation intervals. Limits from other searches and the indirect constraint from measurements of the \PgU and \PZ boson decay widths~\cite{Dobrescu:2014fca} are also shown.}\label{LIM}
\end{figure}

The results of the fit are used to set 95$\%$ confidence level (\CL) upper limits on $\gpq$. Upper limits are computed under a modified frequentist approach, using the \CLs criterion~\cite{CLS1,CLS2}. A profile likelihood ratio is used as the test-statistic and its distribution under the null and alternate hypotheses are determined with asymptotic approximations~\cite{Cowan:2010js}. Limits are shown in Fig.~\ref{LIM} as a function of the resonance mass. Coupling values above the solid curves are excluded at 95\% \CLnp. Systematic uncertainties are treated as nuisance parameters, which are modeled with log-normal priors and profiled in the limit calculations. Values of $\gpq$ greater than 0.3 are excluded at 95\% \CL for the entire mass range. For most of the mass range below 50\GeV, made accessible by the trigger strategy, the exclusion from this analysis is more stringent than the indirect limits set by measurements of the \PZ boson and \PgU meson decay widths~\cite{Dobrescu:2014fca}.

In summary, a search for a low mass \PZpr resonance decaying to \qqbar pairs has been presented, using data from proton-proton collisions at the LHC with a center-of-mass energy of 13\TeV. Jet substructure and decorrelation techniques are implemented to search for narrow resonances over a smoothly falling background of the jet groomed mass. No significant excess is observed above the standard model expectation. Upper limits are placed on the quark coupling strength $\gpq$ of \PZpr bosons with masses between 10 and 125\GeV. Below 50\GeV, the results obtained with this trigger strategy probe the lowest diquark resonance masses reached by a hadron collider.

\begin{acknowledgments}
We congratulate our colleagues in the CERN accelerator departments for the excellent performance of the LHC and thank the technical and administrative staffs at CERN and at other CMS institutes for their contributions to the success of the CMS effort. In addition, we gratefully acknowledge the computing centres and personnel of the Worldwide LHC Computing Grid for delivering so effectively the computing infrastructure essential to our analyses. Finally, we acknowledge the enduring support for the construction and operation of the LHC and the CMS detector provided by the following funding agencies: BMBWF and FWF (Austria); FNRS and FWO (Belgium); CNPq, CAPES, FAPERJ, FAPERGS, and FAPESP (Brazil); MES (Bulgaria); CERN; CAS, MoST, and NSFC (China); COLCIENCIAS (Colombia); MSES and CSF (Croatia); RPF (Cyprus); SENESCYT (Ecuador); MoER, ERC IUT, PUT and ERDF (Estonia); Academy of Finland, MEC, and HIP (Finland); CEA and CNRS/IN2P3 (France); BMBF, DFG, and HGF (Germany); GSRT (Greece); NKFIA (Hungary); DAE and DST (India); IPM (Iran); SFI (Ireland); INFN (Italy); MSIP and NRF (Republic of Korea); MES (Latvia); LAS (Lithuania); MOE and UM (Malaysia); BUAP, CINVESTAV, CONACYT, LNS, SEP, and UASLP-FAI (Mexico); MOS (Montenegro); MBIE (New Zealand); PAEC (Pakistan); MSHE and NSC (Poland); FCT (Portugal); JINR (Dubna); MON, RosAtom, RAS, RFBR, and NRC KI (Russia); MESTD (Serbia); SEIDI, CPAN, PCTI, and FEDER (Spain); MOSTR (Sri Lanka); Swiss Funding Agencies (Switzerland); MST (Taipei); ThEPCenter, IPST, STAR, and NSTDA (Thailand); TUBITAK and TAEK (Turkey); NASU and SFFR (Ukraine); STFC (United Kingdom); DOE and NSF (USA).
\end{acknowledgments}

\bibliography{auto_generated}
\cleardoublepage \appendix\section{The CMS Collaboration \label{app:collab}}\begin{sloppypar}\hyphenpenalty=5000\widowpenalty=500\clubpenalty=5000\vskip\cmsinstskip
\textbf{Yerevan Physics Institute, Yerevan, Armenia}\\*[0pt]
A.M.~Sirunyan$^{\textrm{\dag}}$, A.~Tumasyan
\vskip\cmsinstskip
\textbf{Institut für Hochenergiephysik, Wien, Austria}\\*[0pt]
W.~Adam, F.~Ambrogi, T.~Bergauer, J.~Brandstetter, M.~Dragicevic, J.~Erö, A.~Escalante~Del~Valle, M.~Flechl, R.~Frühwirth\cmsAuthorMark{1}, M.~Jeitler\cmsAuthorMark{1}, N.~Krammer, I.~Krätschmer, D.~Liko, T.~Madlener, I.~Mikulec, N.~Rad, J.~Schieck\cmsAuthorMark{1}, R.~Schöfbeck, M.~Spanring, D.~Spitzbart, W.~Waltenberger, J.~Wittmann, C.-E.~Wulz\cmsAuthorMark{1}, M.~Zarucki
% ...
\vskip\cmsinstskip
\dag: Deceased\\
1:  Also at Vienna University of Technology, Vienna, Austria\\
% ...
\end{sloppypar}
\end{document}